\documentclass[lettersize,journal]{IEEEtran}
\usepackage{amsmath,amsfonts}

\usepackage{cite}
\usepackage{array}

\usepackage{tabularx}
\usepackage{booktabs}

\usepackage{bbding}
\usepackage[caption=false,font=normalsize,labelfont=sf,textfont=sf]{subfig}
\usepackage{multirow}
\usepackage{makecell}
\usepackage{textcomp}
\usepackage{stfloats}
\usepackage{url}
\usepackage{verbatim}
\usepackage{graphicx}
\usepackage[T1]{fontenc} 
\usepackage{amssymb} 
\usepackage{algorithmic}
\usepackage{algorithm}
\usepackage[backref=False]{hyperref}
\def\BibTeX{{\rm B\kern-.05em{\sc i\kern-.025em b}\kern-.08em
    T\kern-.1667em\lower.7ex\hbox{E}\kern-.125emX}}
\usepackage{balance}
\newcommand{\Rmnum}[1]{\uppercase\expandafter{\romannumeral #1}}  
\usepackage{tikz,xcolor,hyperref}

\usepackage{amsmath}
\usepackage{subcaption}
\usepackage{graphicx}
\usepackage{amssymb}

\definecolor{lime}{HTML}{A6CE39}
\DeclareRobustCommand{\orcidicon}{
	\begin{tikzpicture}
		\draw[lime, fill=lime] (0,0)
		circle[radius=0.16]
		node[white]{{\fontfamily{qag}\selectfont \tiny \.{I}D}}; 
	\end{tikzpicture}
	\hspace{-2mm}
}
\foreach \x in {A, ..., Z}{%
	\expandafter\xdef\csname orcid\x\endcsname{\noexpand\href{https://orcid.org/\csname orcidauthor\x\endcsname}{\noexpand\orcidicon}}
}

\begin{document}
\title{Integrated Sensing and Communication for Low-Altitude Security}

\author{Ruixing~Ren\hspace{-1.5mm}\orcidA{},~\IEEEmembership{Graduate~Student~Member,~IEEE},

\thanks{
		
Ruixing Ren is with the School of Electronic and Information Engineering, Beijing Jiaotong University, Beijing 100044, China. (e-mail: renruixing0604@163.com)

		
}
}

\maketitle

\begin{abstract}
  The dense concentration of low-altitude, slow-speed, and small-size targets in the complex low-altitude environment poses significant security challenges, including failures in continuous wide-area sensing and ambiguous target intent, which existing regulatory frameworks struggle to address. Integrated sensing and communication (ISAC), a hallmark of next-generation mobile communication, offers a transformative approach to low-altitude security governance. By leveraging existing cellular infrastructure and spectrum resources, ISAC enables the construction of a seamless wide-area sensing network, supports intelligent feature extraction and intent inference, facilitates real-time collaborative decision-making, and establishes a dynamic trust authentication framework. This article systematically reviews the technical system, analyzes the security challenges, forecasts the enabling value of ISAC, and discusses the resulting open problems and challenges, thereby laying a foundation for future research and industrial implementation.
\end{abstract}


\section{Introduction}
With the completion of the first standard (3GPP Release 18) of 5G-Advanced in 2024, the global evolution of wireless networks toward 6G has formally entered a new phase. Within this technological transition, integrated sensing and communication (ISAC) has emerged as a key enabling technology for both 5G-Advanced and 6G. By enabling the reuse of network infrastructure and spectrum resources, ISAC achieves deep synergy and significant performance enhancement between communication and radar sensing functions \cite{RenTVTIoV}, garnering extensive attention from both academia and industry.

At the same time, the low-altitude economy (LAE), primarily driven by unmanned aerial vehicles (UAVs) and electric vertical take-off and landing aircraft, is becoming a key area globally for cultivating strategic emerging industries and future industries \cite{Jilin}. Defined as a comprehensive economic ecosystem centered on various low-altitude flight activities, which in turn stimulates the integrated development of high-end manufacturing, operational services, and comprehensive support industries, it is widely regarded as a significant representative of new quality productive forces \cite{Waveform, Cooperative}. In major economies worldwide, the development of the LAE has been elevated to strategic importance and is being accelerated through a series of intensive policy measures and industrial plans.

However, the scalable and commercial development of the LAE faces an irreducible prerequisite: airspace security. Compared to traditional medium- and high-altitude aviation, low-altitude airspace exhibits distinctive characteristics, such as aircraft that are low-altitude, slow-speed, and small-size (LSS), complex environments, and dense and highly heterogeneous traffic.  These features render existing regulatory systems relying on discrete sensing methods (e.g. radar and electro-optics technologies) confronted with a series of severe challenges, including extensive coverage blind spots, high construction costs, and difficulties in identifying non-cooperative targets \cite{Challenges}. This lack of effective security governance has become a core bottleneck constraining the efficient utilization of low-altitude resources and the healthy development of the industrial ecosystem.

Against this backdrop, ISAC technology offers a highly promising innovative approach to addressing the challenges of low-altitude security (LAS) governance. Essentially, ISAC aims to upgrade cellular networks from mere communication pipelines to intelligent infrastructure with integrated environmental sensing capabilities \cite{RenUAV,Congcong}. This provides a novel paradigm for tackling the LAS predicament of being undetected, misidentified and uncontrollable: leveraging ubiquitous mobile communication base stations (BSs) as distributed sensing nodes, it is expected to achieve continuous, high-resolution and intelligent situational awareness of low-altitude airspace at an affordable cost, and integrate with communication capabilities to establish a closed-loop security management system covering sensing, cognition and decision-making \cite{Waveform}.

ISAC has emerged as a key technical approach to bolstering the development of the LAE. Current research has been conducted from multiple dimensions. In terms of waveform and signal processing, research focuses on designing waveforms with both high communication reliability and high sensing resolution. For instance, the work of \cite{Waveform} proposed an adaptive waveform, which optimized the bit error rate and sensing sidelobe level by integrating coding and index modulation. The work of \cite{Cooperative} improved the accuracy and robustness of UAV parameter estimation via tensor decomposition and multi-station data fusion.

In network architecture and resource optimization, research focuses on improving the overall system efficiency via collaborative and intelligent technologies. Ref. \cite{11205853} introduced a hybrid reconfigurable intelligent surface to compensate for the elevation angle limitation of BS antennas and improved the sensing signal-to-noise ratio for high-altitude targets. Ref. \cite{10879807} balanced sensing and communication performance through the joint design of collaborative beamforming and UAV trajectories.

Several studies propose novel system frameworks and performance metrics. Ref. \cite{11315846} presentd a full-duplex ISAC system prototype, enabling simultaneous co-frequency sensing and communication via interference cancellation. Ref. \cite{11151696} modeled low-altitude surveillance as a compressed sensing imaging problem and introduces physical embedding learning to mitigate off-grid errors. Ref. \cite{10950390} innovatively proposed the metric of sensing capacity to evaluate the upper limit of BSs' ability to detect multiple targets simultaneously.

These efforts have advanced the technological development of ISAC in low-altitude environments and preliminarily verified its potential to improve sensing accuracy, communication efficiency and resource utilization. However, most existing studies focus on enhancing the communication-sensing performance of ISAC systems themselves or address specific technical challenges (e.g., interference management, trajectory design). Although several studies mention the importance of LAE \cite{Challenges,11207184}, their analyses are often embedded in the framework of ISAC performance optimization, lacking a comprehensive sorting and in-depth analysis of security challenges, ISAC enabling paths and existing problems from the systematic perspective of LAE governance.

This article thus focuses on the ISAC technical system for low-altitude security.  First, Section \ref{2} analyzes the unique security challenges of low-altitude scenarios. Section \ref{3} then prospects ISAC-enabled LAE for each challenge. Section \ref{4} discusses the current open problems and challenges. Section \ref{5} concludes the article. To the authors' knowledge, this is the first review article focusing on ISAC for LAS.

\begin{figure*}[htbp]
	\centerline{\includegraphics[width=6.9in,keepaspectratio]{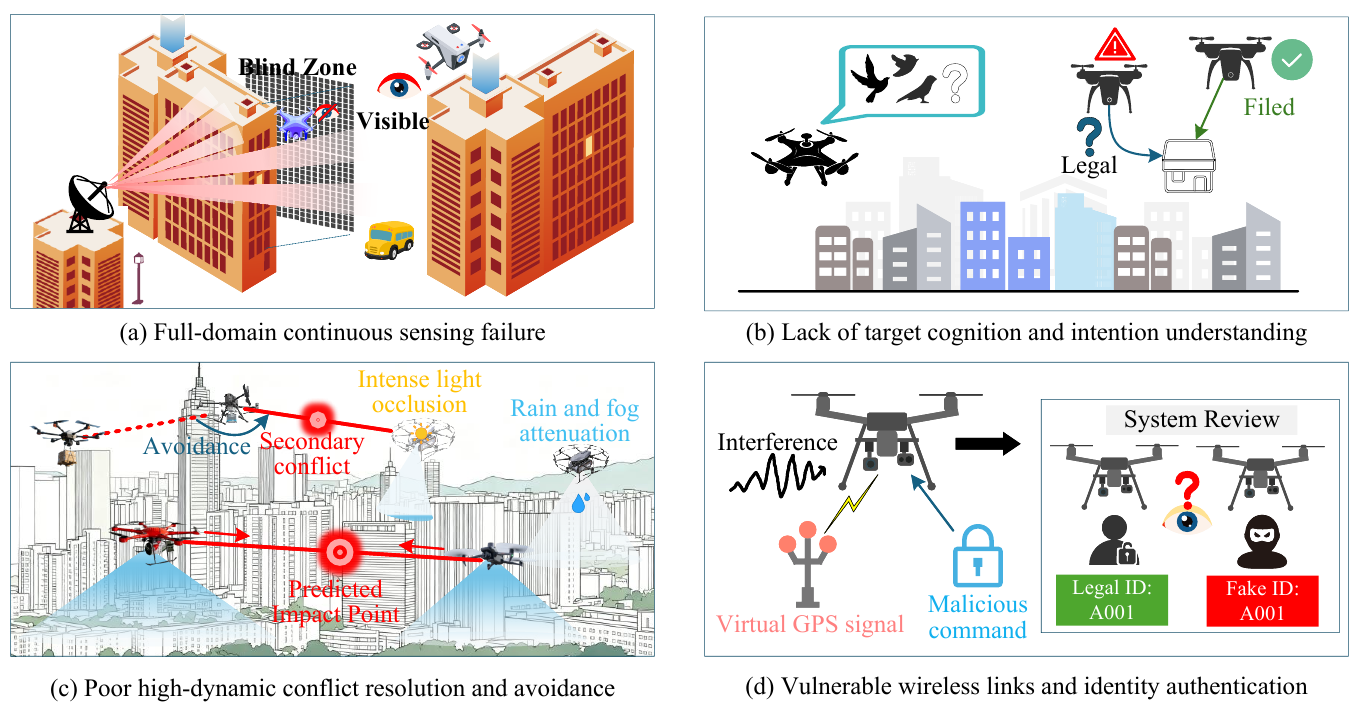}}
	\caption{Key Challenges in LAS}
	\label{fig1}
\end{figure*}

\section{Security Analysis of Low-Altitude Scenarios} \label{2}
With the vigorous development of the LAE, its airspace environment is fundamentally different from that of traditional medium- and high-altitude civil aviation and ground transportation. The unique physical and motion characteristics of aircraft in this scenario constitute fundamental challenges for its security management.

\subsection{Physical and Motion Characteristics of Low-Altitude Targets}
Low-altitude aircraft, especially dominant UAVs, generally exhibit the following characteristics \cite{Waveform,FeifeiGao}:
\begin{itemize}
	\item Low altitude: Operating typically below 1,000 meters, they often traverse complex geographical environments composed of buildings, mountains and trees, with signals prone to obstruction.
	\item Low speed: Featuring low typical cruising speeds (e.g., multi-rotor UAVs usually range from 10 to 50 km/h) and the capability of long-duration stationary hovering.
	\item Small size: With small radar cross-section and weak optical signatures, they have high stealth against traditional high-power long-range detection methods.
	\item High maneuverability: Capable of vertical takeoff and landing, rapid speed change, emergency hovering and flexible turning, with nonlinear and unpredictable motion patterns.
	\item Large-scale/heterogeneity: Various aircraft (consumer-grade, industrial-grade, manned, etc.) will coexist densely in future airspace, differing in performance, purpose and communication protocols.
\end{itemize}

\subsection{Key Security Challenges}
These characteristics collectively give rise to the following core security challenges that must be addressed for large-scale low-altitude applications.

\noindent\textit{(1) Full-Domain Continuous Sensing Failure}

As shown in Fig. \ref{fig1}(a), the low-altitude environment has extensive obstructions caused by terrain and ground objects, where traditional radars form numerous blind spots, leading to a sharp decline in detection performance. Optical monitoring is severely constrained by adverse weather and lighting conditions such as nighttime, haze, rain and snow \cite{RenTVTIoV}. This results in a systematic sensing gap of undetected or lost tracking for LSS targets. Targets are prone to disappearing among urban buildings, making it difficult for air traffic control systems to generate and maintain a real-time, continuous, blind-spot-free comprehensive air situation map, with numerous unknown targets posing potential risk sources.

\noindent\textit{(2) Lack of Target Cognition and Intention Understanding}

Obtaining only target position and velocity vectors constitutes preliminary information, which is far from sufficient for effective security assessment and threat classification. As shown in Fig. \ref{fig1}(b), existing technical means struggle to reliably distinguish morphologically similar UAVs, birds, or aerial floating objects at long distances \cite{FeifeiGao}. 

More importantly, systems cannot distinguish whether a flight activity is an approved legal operation, unintentional unreported flight (grey flight), or malicious unauthorized intrusion (black flight). They lack the capability of real-time analysis based on multi-dimensional data for high-risk behavioral patterns (e.g., abnormal loitering and formation gathering of targets around sensitive areas \cite{ZhaoPlatooning}), failing to achieve the cognitive leap from trajectory tracking to intention understanding and threat determination.

\noindent\textit{(3) Poor High-Dynamic Conflict Resolution and Avoidance}

As shown in Fig. \ref{fig1}(c), low-altitude airspace, especially urban airspace, is a canyon densely populated with static obstacles (e.g., power lines, communication towers, and high-rise buildings). Meanwhile, a dense dynamic traffic flow composed of numerous aircraft will exist in the future. Currently, on-board obstacle avoidance sensors (e.g., visual, ultrasonic, and LiDAR) have limited operating ranges and unstable performance under harsh environments such as strong light and smoke. The existing system lacks an air traffic management infrastructure with wide-area coverage, low-latency communication support, and collaborative decision-making capabilities, failing to provide real-time conflict prediction and automated/assisted cooperative resolution commands for high-density, heterogeneous aircraft swarms, resulting in significant large-scale collision risks \cite{Resolution}.

\noindent\textit{(4) Vulnerable Wireless Links and Identity Authentication}

The command and control, data transmission, and navigation positioning of low-altitude aircraft are highly dependent on radio data links and global satellite navigation systems (GNSS). As shown in Fig. \ref{fig1}(d), these wireless links are vulnerable to intentional or unintentional interference, signal spoofing, and even remote hijacking in complex electromagnetic environments, resulting in aircraft loss of control, deviation, or data leakage. Meanwhile, the existing regulatory system lacks a robust identity authentication mechanism that uniquely and tamper-proof binds an aircraft’s digital identity (e.g., broadcast identification code) to its physical entity. This enables illegal intruders to easily falsify identities, making their malicious activities difficult to block in real time and trace afterward, creating significant security and legal loopholes.

\section{ISAC-Enabled LAS} \label{3}
Addressing the security challenges enumerated in Section \ref{2}, this section elaborates on the approaches and advantages of ISAC in empowering LAS.

\begin{figure}[t]
	\centerline{\includegraphics[width=3.6in,keepaspectratio]{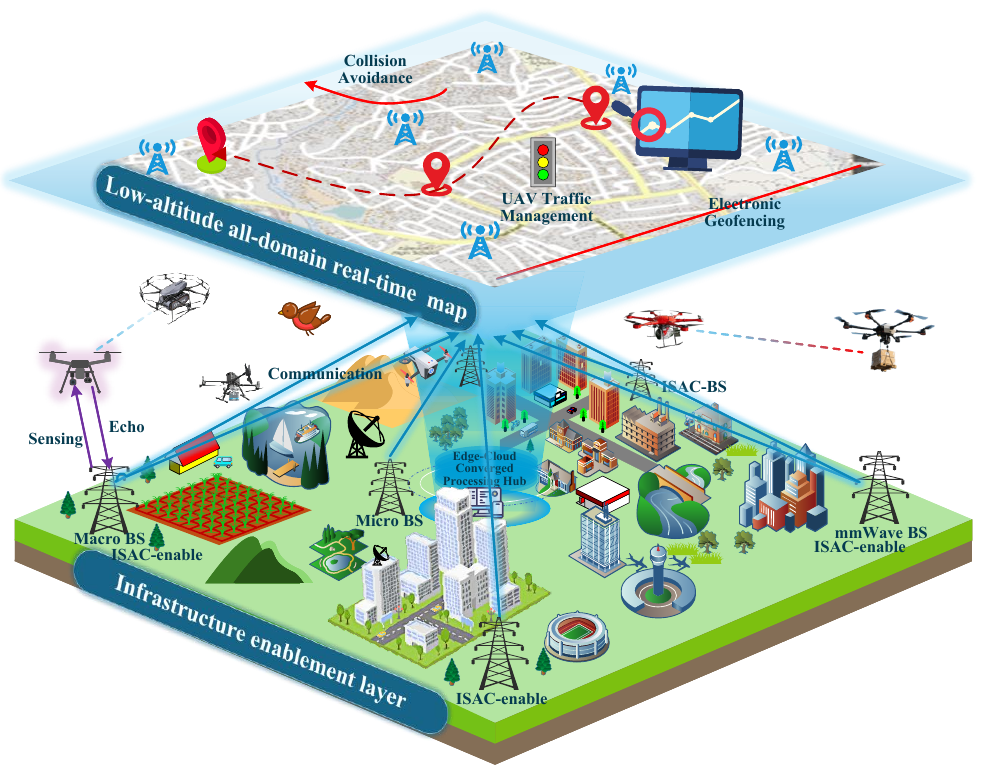}}
	\caption{Wide-Area Continuous Seamless Sensing Network}
	\label{fig2}
\end{figure}
\subsection{Wide-Area Continuous Seamless Sensing Network}
To systematically address the challenge of full-domain continuous sensing failure in low-altitude airspace, this article proposes a promising solution: upgrading widely distributed cellular communication networks into a continuous seamless sensing network. As illustrated in Fig. \ref{fig2}, ISAC technology enables existing networks to form a blind-spot-free, continuous detection sensing network covering the entire low-altitude airspace while fulfilling their communication functions, thereby achieving real-time and transparent control of airspace situational awareness \cite{RenUAV}.

At the infrastructure level, widely deployed macro BSs, micro BSs, and future millimeter-wave BSs form a natural dense sensing node network. Their sites are usually carefully planned and located at urban commanding heights or key coverage areas, laying a physical foundation for establishing an unobstructed sensing field of view. By upgrading existing BSs with software-defined radio and enhancing antenna array hardware, they can transmit or reuse optimized sensing waveforms while emitting communication signals, and receive reflected signals from aerial targets with high sensitivity \cite{RenRIS}. This fully leverages existing infrastructure, avoiding the substantial costs and deployment cycles associated with building new dedicated radar stations. Meanwhile, the system operates entirely within licensed and compliant mobile communication frequency bands and power levels, ensuring public electromagnetic safety and environmental friendliness.

At the signal and information processing level, by designing ISAC waveforms with excellent autocorrelation and cross-correlation characteristics, and employing advanced multiple-input multiple-output beamforming and signal processing algorithms, multi-dimensional information (e.g., range, velocity, and angle) of LSS targets can be extracted in real time from complex environments with strong clutter and interference \cite{Waveform}. Although the sensing information of a single BS is limited, relying on low-latency mobile backhaul networks, local data acquired by multiple BSs undergoes spatiotemporal registration, data association, and trajectory fusion at the edge or cloud, enabling the synthesis of a wide-coverage, continuous, and unified real-time airspace situational map.

\begin{figure}[t]
	\centerline{\includegraphics[width=3.1in,keepaspectratio]{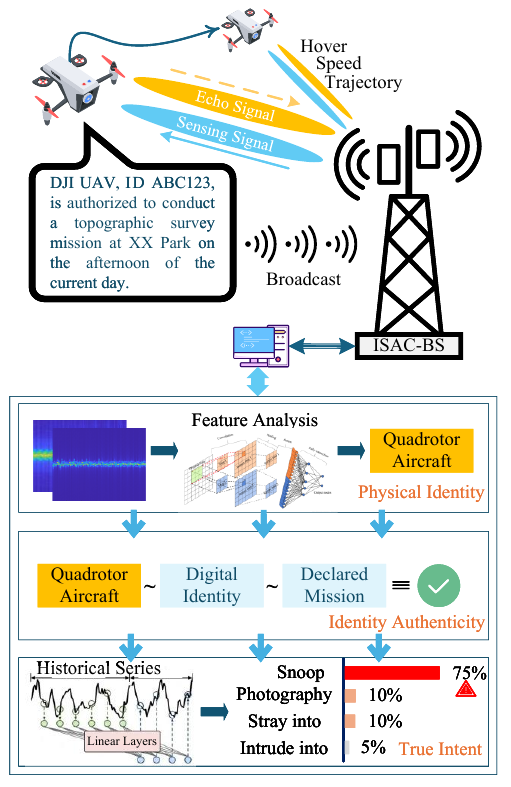}}
	\caption{Intelligent Feature Extraction and Intention Understanding}
	\label{fig3}
\end{figure}
\subsection{Intelligent Feature Extraction and Intention Understanding}
To address the cognitive dilemma of ambiguous identity and unclear intention of low-altitude targets, an ISAC-based intelligent cognitive enhancement framework needs to be established. As shown in Fig. \ref{fig3}, ubiquitous signals in communication networks are converted into sensing sources for in-depth target understanding, and through multi-level information fusion and intelligent analysis, the cognitive leap from raw data to threat determination is achieved.

At first, raw channel state information and radar echoes captured by ISAC BSs contain the unique physical fingerprints of targets. By deploying lightweight AI models for real-time analysis of signal features such as micro-Doppler and high-resolution range profiles, feature vectors reflecting targets’ mechanical structures (e.g., number of rotors, rotational speed), approximate size, and surface materials can be extracted in a non-cooperative manner \cite{FeifeiGao}. This provides a quantifiable basis for distinguishing between consumer-grade UAVs, industrial-grade UAVs, birds, and aerial floating objects.

Then, the system performs spatiotemporal correlation and fusion of the aforementioned physical features with accessible signaling data in cyberspace (e.g., temporary device identifiers, signal strength history) and planned information (e.g., declared flight plans, electronic fences). For instance, it matches and verifies the detected physical features of UAVs with their broadcast remote radio identifiers, or compares their trajectories with approved air routes. This enables cross-validation of targets’ physical entities, digital identities, and claimed missions.

For high-level intention understanding, it is necessary to construct a behavior analysis engine based on temporal deep learning models or knowledge graphs to model targets’ continuous trajectories, speed changes, and regional dwell patterns \cite{ZhaoTrajectory}. Integrating context provided by geographic information systems (e.g., distribution of sensitive facilities, airspace structure), the system can identify abnormal behavioral patterns such as loitering, approaching reconnaissance, and formation gathering, and perform probabilistic prediction of potential intentions (e.g., straying, snooping, malicious intrusion) and threat level assessment.

\begin{figure*}[t]
	\centerline{\includegraphics[width=6.6in,keepaspectratio]{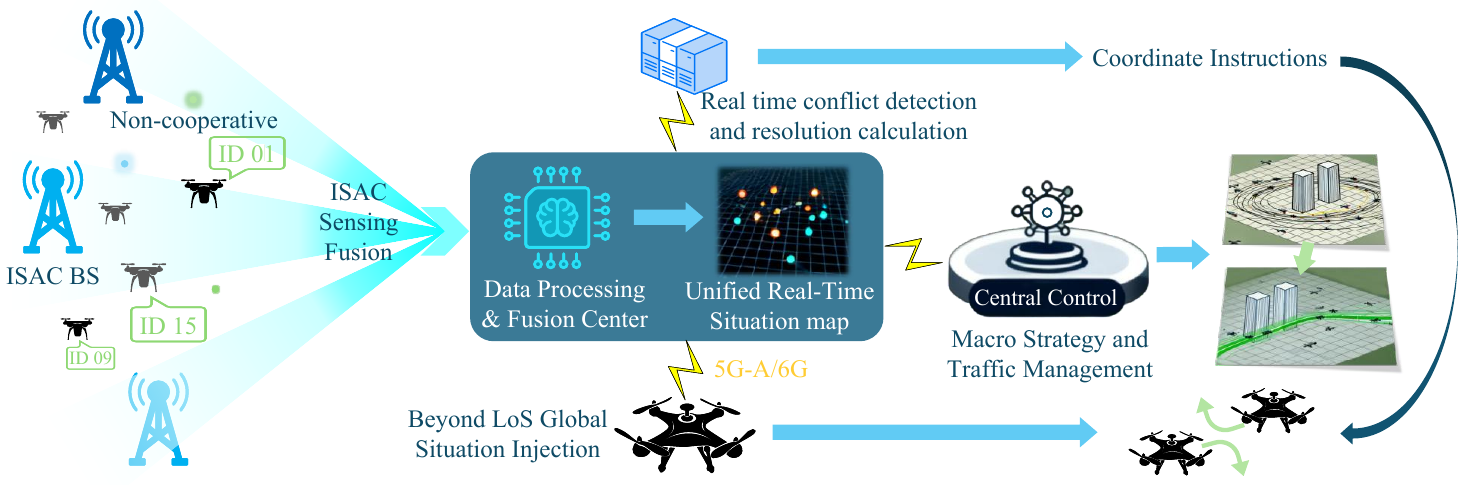}}
	\caption{Cooperative Decision-Making Architecture Based on ISAC Unified Airspace Situational Map}
	\label{fig4}
\end{figure*}

\subsection{Real-Time Collaborative Situation Sharing and Decision Support}
To address the conflict resolution challenge, as shown in Fig. \ref{fig4}, the constructed ISAC network can generate a full-domain real-time situational map, which is converted into shareable public information products via low-latency communication, thereby promoting the establishment of a new security paradigm driven by network-wide perception and aerial distributed collaboration \cite{Cooperative,ZhaoUAVMEC}.

The implementation of this architecture is rooted in a design logic where the three loops of sensing, communication, and control are tightly coupled.
In the sensing and fusion loop, through multi-BS collaboration, the ISAC network fuses the trajectories of all cooperative targets and non-cooperative targets within its coverage, generating a four-dimensional spatiotemporal situational map with a single information source and centimeter-level accuracy, including position, velocity, heading, and timestamp \cite{Feifei}. This helps resolve the issues of inconsistent, incomplete, and ambiguous situational awareness among aircraft in traditional modes.

In the communication and distribution loop, leveraging the inherent low-latency, high-reliability, and network slicing capabilities of 5G-A/6G networks, the aforementioned unified situational map is delivered as a critical data service to three types of nodes on demand, securely, and in real time \cite{RenITS}: first, aerial aircraft, providing beyond-visual-line-of-sight global environmental awareness for their autonomous driving systems; second, edge computing nodes, used to execute fast conflict detection and resolution algorithms; third, central management and control platforms, for macro airspace traffic management and strategy formulation. This ensures the unimpeded flow of critical security information within the system.

In the decision-making and control loop, each node conducts collaborative decision-making based on a fully consistent situational map. Aircraft perform autonomous conflict prediction and route replanning in accordance with preset rules; edge nodes coordinate multiple aircraft for distributed resolution; central platforms dynamically adjust airspace structure or issue traffic instructions. Thus, the entire system evolves from a loose sensing-independent response mode to an integrated sensing-sharing-collaborative decision-making system.

\begin{figure}[t]
	\centerline{\includegraphics[width=3.6in,keepaspectratio]{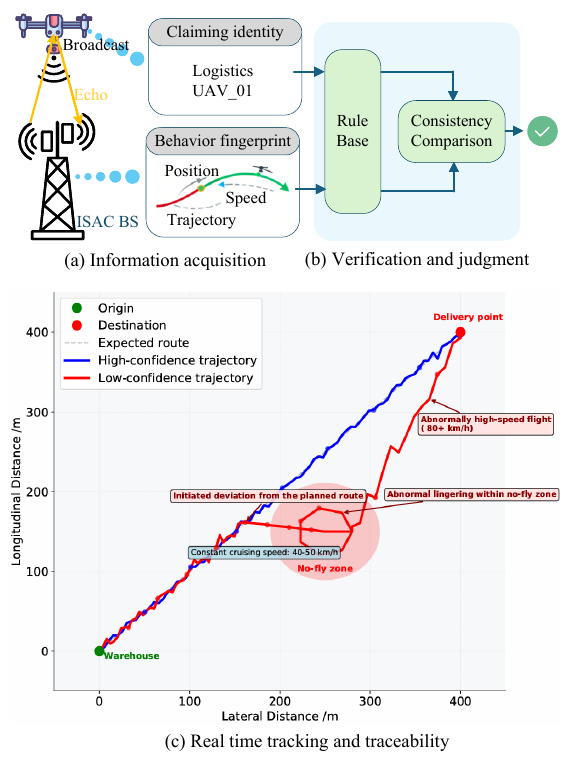}}
	\caption{ISAC-Based Dynamic Trusted Authentication System}
	\label{fig5}
\end{figure}

\subsection{ISAC-Based Dynamic Trusted Authentication System}
To address the core security challenges posed by vulnerable low-altitude wireless links and identity spoofing, an ISAC-based cross-layer trusted authentication system is a promising solution. Compared with traditional solutions, this system is no longer limited to identity verification only at the communication protocol layer. Instead, it takes the physical layer behavioral features provided by ISAC as endogenous security elements, enabling continuous two-way verification between communication digital identities and physical entity behaviors \cite{Yaoyu}.

As shown in Fig. \ref{fig5}(a), the system first captures two types of key information in parallel. On the one hand, it obtains the target’s claimed identity by parsing communication protocols, such as broadcast remote identification codes and temporary identifiers for network access. On the other hand, leveraging the sensing capability of the ISAC network, it independently acquires the physical behavioral fingerprint of the signal transmitter, including precise real-time position, 3D motion trajectory, velocity variation pattern, and signal angle of arrival, etc. These two types of evidence are synchronously correlated in time and space, forming the basis for comparison.

Based on the collected multi-source information, as shown in Fig. \ref{fig5}(b), the system performs dynamic verification of the target identity through a built-in claimed-behavior consistency rule base. For example, Rule 1: If a signal claims to belong to logistics UAV 01, its motion trajectory should conform to a continuous path from the warehouse to the delivery point, and its speed should be within the cruising range of this model. Rule 2: The geographic location of its signal transmitter should be consistent with the broadcast GNSS position within the error tolerance. The system conducts risk assessment by real-time comparing the behavioral expectations implied by the claimed identity with the actual behaviors perceived by ISAC. Any significant deviation (e.g., appearance in a no-fly zone, sudden trajectory change, inconsistency between identity and UAV model characteristics) will trigger a trust score decay or an abnormal alarm.

Finally, the system dynamically evaluates and maintains an integrated trust credential for each aerial entity. This credential not only includes its digital identity but also binds its continuously verified historical behavioral features. As shown in Fig. \ref{fig5}(c), the system can track the target’s motion trajectory in real time, distinguish between high-confidence and low-confidence paths, and visually mark and record abnormal behaviors. In the event of malicious behaviors (e.g., signal spoofing, intrusion attacks), the system can provide non-repudiable audit trails, accurately trace to specific physical devices and their behavioral histories, and realize the transformation from anonymous signals to accountable entities.

\section{Open Problems and Challenges} \label{4}
Although ISAC technology offers the revolutionary solutions for LAS governance, it still faces multiple core challenges in advancing from theory to large-scale implementation.

\subsection{Sensing Performance Bottlenecks in Complex Environments}
LSS targets feature small radar cross-sections and weak signal reflections and they are also susceptible to building occlusion, terrain clutter, electromagnetic interference and severe weather, resulting in extremely low signal-to-clutter ratios of perceived signals. Existing multi-BS data fusion algorithms suffer from cumulative spatiotemporal registration errors and difficulties in heterogeneous data consistency verification, making it hard to achieve high-precision full-domain situational awareness without blind spots. Meanwhile, the passive detection capability for non-cooperative targets is insufficient; extracting stable physical features and motion fingerprints without relying on active target cooperation remains a technical challenge.

\subsection{Insufficient Generalization of Intelligent Cognition and Intent Understanding}
Low-altitude targets exhibit high heterogeneity, including consumer-grade, industrial-grade, manned aircraft and birds. Their behavioral patterns are complex with strong randomness and antagonism (e.g., malicious targets deliberately camouflage their trajectories). Existing AI models mostly rely on labeled data training in specific scenarios, with limited generalization capability, making them difficult to adapt to dynamically changing low-altitude environments. Moreover, intent understanding requires the fusion of multi-dimensional heterogeneous data (physical features, signaling data, geographic information, etc.). The contradiction between data privacy protection and cross-domain data sharing is prominent, and realizing data value mining under compliance premises remains the key constraint on the upgrading of cognitive capabilities.

\subsection{Real-Time Guarantee for Low-Latency Collaborative Decision-Making}
In high-dynamic low-altitude traffic scenarios, conflict resolution and path replanning require millisecond-level latency. Yet end-to-end latency of multi-BS sensing data transmission, edge node collaborative computing and command delivery is susceptible to network congestion. Existing distributed decision-making architectures lack a balance between global optimality and local autonomy, hindering efficient collaborative obstacle avoidance for high-density aircraft swarms. Establishing a low-latency, high-reliability collaborative decision-making mechanism under limited communication resource constraints to ensure command consistency and execution safety constitutes a key prerequisite for supporting large-scale low-altitude operations.

\subsection{Robustness Challenges of Security Authentication in Adversarial Environments}
Low-altitude wireless links are vulnerable to malicious attacks such as signal spoofing, jamming and hijacking, and existing dynamic trusted authentication systems cannot handle high-intensity adversarial scenarios. For instance, attackers can forge physical behavior fingerprints or tamper with GNSS positioning information to evade claim-behavior consistency verification. Meanwhile, the lack of cross-operator and cross-airspace authentication interoperability mechanisms leads to the breakage of the full-domain trust chain. Designing anti-counterfeiting and non-repudiation physical layer security features and establishing a distributed, lightweight cross-domain trusted authentication network constitute the core requirement for ensuring system security.

\section{Conclusion} \label{5}
This article focused on the deep integration of LAS and ISAC technology, conducting a systematic analysis around challenges, empowerment and prospects. First, it clarified the uniqueness of the LAE, which gives rise to four core security bottlenecks. Second, it elaborated on the empowerment pathways of ISAC technology. Finally, it identified the key challenges in the current implementation of the technology. As the first overview of ISAC technology in the field of LAE, this article only presented a partial research landscape. However, it is hoped that the relevant discussions can stimulate academic interest and in-depth exploration in the future evolution of LAS.

\bibliographystyle{ieeetr} 
\bibliography{MyRefs} 
~~~\\
~~~\\

\end{document}